\documentclass[aap,MSNbibl,seceqn,citesort,dvips]{arximspdf}

\doi{10.1214/12-AAP850} 
\volume{23}
\issue{2}
\pubyear{2013}
\firstpage{665}
\lastpage{692}

\makeatletter

\newtheorem{theorem}{Theorem}[section]
\newtheorem{lemma}[theorem]{Lemma}
\newtheorem{proposition}[theorem]{Proposition}

\newtheorem{corollary}[theorem]{Corollary}

\newproclaim{definition}[theorem]{Definition}
\newproclaim{example}[theorem]{Example}

\newproclaim{assumption}[theorem]{Assumption}

\newproclaim{remark}[theorem]{Remark}

\newtheorem{claim}[theorem]{Claim}

\newcommand{\argmax}{\mathop{\arg\max}}
\newcommand{\argmin}{\mathop{\arg\min}}
\newcommand{\esssup}{\mathop{\operatorname{ess}\sup}}
\newcommand{\essinf}{\mathop{\operatorname{ess}\inf}}
\newcommand{\epi}{\operatorname{epi}}

\newcommand{\conv}{\operatorname{conv}}

\newcommand{\sq}[2][n]{\{#2\}_{#1\in\N}}

\newcommand{\R}{{\mathbb R}}
\newcommand{\N}{{\mathbb N}}
\newcommand{\PP}{{\mathbb P}}
\newcommand{\QQ}{{\mathbb Q}}
\newcommand{\hQ}{\hat{\QQ}}
\newcommand{\EE}{{\mathbb E}}
\newcommand{\FF}{{\mathcal F}}
\newcommand{\UU}{\mathbb{U}}
\newcommand{\VV}{\mathbb{V}}
\newcommand{\DD}{{\mathcal D}}
\newcommand{\MM}{{\mathcal M}}
\newcommand{\NN}{{\mathcal N}}
\newcommand{\FFF}{{\mathbb F}}
\newcommand{\CC}{{\mathcal C}}
\newcommand{\EN}{{\mathcal E}}
\newcommand{\XX}{{\mathcal X}}
\renewcommand{\AA}{{\mathcal A}}
\newcommand{\ba}{{\mathrm{ba}}}
\newcommand{\eps}{\varepsilon}
\newcommand{\vp}{\varphi}

\newcommand{\el}{{\mathbb L}}
\newcommand{\lzer}{\el^0}
\newcommand{\lone}{\el^1}
\newcommand{\lpee}{\el^p}
\newcommand{\linf}{\el^{\infty}}
\newcommand{\linfd}{(\linf)^*}

\newcommand{\bsx}{{\mathbf{x}}}
\newcommand{\bsy}{{\mathbf{y}}}
\newcommand{\sA}{{\mathcal A}}
\newcommand{\sC}{{\mathcal C}}
\newcommand{\sF}{{\mathcal F}}
\newcommand{\hH}{{\hat{H}}}
\newcommand{\sK}{{\mathcal K}}
\newcommand{\sP}{{\mathcal P}}
\newcommand{\sS}{{\mathcal S}}
\newcommand{\CR}{2^{\R^d}_{c}}
\newcommand{\A}{\sA}
\newcommand{\Xc}{\XX^c}
\newcommand{\ac}{\alpha_{\sC}}
\newcommand{\bn}{B^{N}}
\newcommand{\sn}{S^{N}}
\newcommand{\sba}{S^{\ba}}
\newcommand{\sbap}{\sba_+(1)}
\newcommand{\sbayp}{\sba_+(y)}
\renewcommand{\sl}{S^{\lone}}
\newcommand{\slp}{\sl_+(1)}
\newcommand{\slyp}{\sl_+(y)}
\newcommand{\sey}{\sba_{\eps}(y)}
\newcommand{\sley}{\sl_{\eps}(y)}
\newcommand{\vba}{v^{\ba}}
\newcommand{\hQy}{\hat{\QQ}^{(y)}}
\newcommand{\hUU}{\hat{\UU}}
\newcommand{\fx}{f^{(x)}}
\newcommand{\Rinf}{\R\cup\{+\infty\}}
\newcommand{\Hx}{H^{(x)}}
\newcommand{\ux}{\underline{x}}

\makeatother

\begin{document}
\begin{frontmatter}

\title{On utility maximization under convex portfolio~constraints}
\runtitle{Utility maximization under constraints}

\begin{aug}
\author[A]{\fnms{Kasper} \snm{Larsen}\ead[label=e1]{kasperl@andrew.cmu.edu}}
\and
\author[B]{\fnms{Gordan} \snm{\v{Z}itkovi\'{c}}\corref{}\thanksref{t2}\ead[label=e2]{gordanz@math.utexas.edu}}
\pdfauthor{Kasper Larsen, Gordan Zitkovic}
\runauthor{K. Larsen and G. \v Zitkovi\'{c}}
\affiliation{Carnegie Mellon University and University of Texas at Austin}
\address[A]{Department of Mathematical Sciences\\
Carnegie Mellon University\\
Pittsburgh, Pennsylvania \\
USA\\
\printead{e1}}
\address[B]{Department of Mathematics\\
University of Texas at Austin\\
Austin, Texas\\
USA\\
\printead{e2}}
\end{aug}

\thankstext{t2}{Supported in part by NSF Grant DMS-09-55614 during the preparation of this work.
Any opinions, findings and conclusions
or recommendations expressed in this material are those of the
author and do not necessarily reflect those of the
National Science Foundation.}

\received{\smonth{2} \syear{2011}}
\revised{\smonth{1} \syear{2012}}

%
\begin{abstract}
We consider a utility-maximization problem in a general semimartingale
financial model, subject to constraints on the number of shares held in
each risky asset. These constraints are modeled by predictable
convex-set-valued processes whose values do not necessarily contain the
origin; that is, it may be inadmissible for an investor to hold no
risky investment at all. Such a setup subsumes the classical
constrained utility-maximization problem, as well as the problem where
illiquid assets or a random endowment are present.

Our main result establishes the existence of optimal trading strategies
in such models under no smoothness requirements on the utility
function. The result also shows that, up to attainment, the dual
optimization problem can be posed over a set of countably-additive
probability measures, thus eschewing the need for the usual
finitely-additive enlargement.
\end{abstract}

%

\end{frontmatter}

\section{Introduction and notation}\label{intro}
\subsection{The existing literature}
The study of utility maximization in con\-tinuous-time stochastic
models of financial markets dates back to the seminal contributions of
Robert Merton~\cite{Mer69,Mer71}. General complete Brownian models were
considered by Karatzas, Lehoczky and Shreve~\cite{KarLehShr87} and Cox and
Huang~\cite{CoxHua89}, where the authors used convex-analytic (duality)
techniques to characterize the optimizer. Duality techniques for
incomplete It\^{o}-process models were first developed by Karatzas et
al.~\cite{KarLehShrXu91}, and in a general
semimartingale setting, by Kramkov and Schachermayer~\cite{KraSch99,KraSch03}.

Cvitani\'{c} and Karatzas~\cite{CviKar92} extended the existence
results of
Karatzas et al.~\cite{KarLehShrXu91} to incorporate convex constraints on
the fraction of wealth invested in the risky securities. In the same
It\^{o}-process driven setting, Cuoco~\cite{Cuo97} attacked the primal problem
directly and established the existence of optimizers when investors face
convex constraints either on the number of shares or on the amount
invested.

Relying on a version of the optional decomposition theorem of
F\"ollmer and Kramkov~\cite{FolKra97}, Pham~\cite{Pha02} and
Mnif and Pham~\cite{MniPha01} studied constrained optimization in the
general semimartingale setting. In~\cite{Pha02}, the author
generalized the shortfall objective considered in F\"ollmer and
Leukert~\cite{FolLeu00}, while, in~\cite{MniPha01},
investors, subject to either convex constraints on the number of risky
securities, or American-type constraints on the wealth process, have
been considered.
As in~\cite{KraSch03}, both~\cite{Pha02} and~\cite{MniPha01} used
``Koml\'os-type'' arguments to establish the existence of primal
optimizers. The question of dual existence was, however, left open (see
the discussions on page 154 in~\cite{Pha02} and page 167 in
\cite{MniPha01}). Constraints on the fractions of wealth
invested in the risky securities were investigated in~\cite{Lon04} by
Long who established the existence of optimizers under a number of
strong additional assumptions.

Among several authors who studied the existence of optimizers for
nonsmooth utility functions, we mention Bouchard, Touzi and Zeghal
\cite{BouTouZeg04}, and we direct the reader to consult their references.
The recent counterexample of Westray and Zheng~\cite{WesZhe11}
illustrates some of the counterintuitive phenomena nonsmooth dual
objectives can produce.

\subsection{Our contributions} The analysis in most of the papers mentioned
above requires that the investor be allowed to choose not to invest in the
risky securities at all, with~\cite{MniPha01} serving as a notable
exception. In the present paper, no such condition is imposed:
one might be forced to invest in risky assets some or all of the time;
the idea to apply constraints not containing the origin to
utility-maximization problems goes back, at least, to the work~\cite{Kal99}
of Kallsen; see also~\cite{Kal02}. The study of
such a general class of constraints is interesting from both
mathematical and economical points of view. Mathematically, this setup
produces an interesting convex-analytic situation where
the support
function is no longer necessarily nonnegative. Economically, such
constraints
correspond to the case when some of the available
assets are not perfectly liquid and the investor is effectively forced to
hold them. The case of a terminal random endowment, studied by
Cvitani\'c, Schachermayer and Wang~\cite{CviSchWan01} and Hugonnier and Kramkov
\cite{HugKra04} among others, can be embedded in our setting---it
corresponds to a constraint which forces the investor to hold one unit
of a
specific asset to maturity. Finally, a number of classical constraints,
including the prohibition or restriction of short selling, can be
interpreted as convex portfolio constraints, and fit into our framework.

There are two main results in this paper and they both apply to a
general semimartingale model of a financial market. The first
one\vadjust{\goodbreak}
establishes the existence of the primal and dual optimizers in the
constrained utility-maximization problem, with the dual
problem defined over a class of finitely-additive measures. The
conjugacy of the primal and the dual value functions is an integral
part of
our result. The only assumption imposed on the utility function,
besides the defining properties of concavity, monotonicity and the
Inada condition at zero, is the reasonable asymptotic elasticity of
\cite{KraSch99}.

Our second result is that the finitely-additive relaxation is, up to
attainment, in fact, not necessary, and that the dual problem can be
posed over a class of countably-additive measures. This result
generalizes Theorem~2.2(iv) of~\cite{KraSch99} to our constrained
case; in particular, it subsumes the case of an unspanned endowment
considered in~\cite{CviSchWan01}. The main technical difficulty we had
to overcome is the absence of semicontinuity in the appropriate
direction of the dual objective function (in general, this objective
is not upper semicontinuous). Our solution is based on
Theorem 2.2(iv) of~\cite{KraSch99} and methods of locally-convex convex analysis.
This countably-additive relaxation has several practical implications.
First of all, the classical stochastic-optimal-control framework and
the corresponding tools and notions, such as the dynamic programming
principle and the associated Hamilton--Jacobi--Bellman equation, rely on
having stochastic processes (in our case, densities of
countably-additive measures)
as controls. These tools are not immediately available or applicable
in more general settings (such as the finitely-additive one).
Furthermore, the existence of $\eps$-optimal countably-additive
measures
serves as a first step toward an efficient numerical treatment of the
problem.

As far as no-arbitrage-type assumptions are concerned, our main
existence and conjugacy results are provided under the abstract
assumption of closedness and boundedness in probability (convex
compactness in the language of \v Zitkovi\'{c}~\cite{Zit09}) of the
$\lzer_+$-solid hull $\CC(x)$ of the set of terminal wealths of
admissible portfolios with
initial wealth $x$. This condition is weaker than the celebrated
No Free Lunch with Vanishing Risk (NFLVR) of Delbaen and Schachermayer
\cite{DelSch94} and is reminiscent of the No Unbounded Profit with
Bounded Risk (NUPBR) condition of Karatzas and Kardaras \cite
{KarKar07}. Indeed, given the presence of constraints, the classical
NFLVR can be too strong, as the
constraints will often prevent the investor from making riskless profit,
even if the asset prices would admit arbitrage in the unconstrained market.
Using a new closedness result of Czichowsky
and Schweizer~\cite{CziSch11} for sets of constrained stochastic integrals
in the semimartingale topology, we give a general and easy-to-check
sufficient condition for the convex compactness of~$\CC(x)$.

\subsection{Notation and function spaces}
All stochastic objects are defined on a filtered probability space
$(\Omega,\sF,(\sF_t)_{t\in[0,T]},\PP)$ where the $T\in(0,\infty
)$ is
the time
horizon, and the underlying filtration $\FFF:= (\sF_t)_{t\in[0,T]}$
satisfies the usual conditions. For $p\in(0,\infty]$, $\lpee$ denotes
the Lebesgue space $\lpee(\Omega,\sF,\PP)$, and $\lzer$ denotes the
collection of all $\PP$-a.s. equivalence classes of finite-valued
random \mbox{variables} on $(\Omega,\FF)$ (topologized by convergence in
probability). If not stated otherwise, all processes are assumed to be
c\`{a}dl\`{a}g and $\FFF$-adapted, with the exception of processes which
serve as integrands in stochastic integrals; those are always assumed
to be $\FFF$-predictable.

While none of our results require their mention in the
statements, finitely-additive measures are used quite frequently in
proofs. We naturally identify finite-valued finitely-additive set
functions on
$(\Omega,\FF)$ which vanish on $\PP$-null events with the topological
dual $\ba:=\linfd$ of $\linf$; see~\cite{RaoRao83} for further
details. The dual pairing of $\ba$ and $\linf$ is denoted
by $\langle\cdot,\cdot\rangle\dvtx\ba\times\linf\to\R$ and the
(dual) norm
${\|\cdot\|}$ on $\ba$ is given by
${\|\QQ\|}:=\sup\{|\langle\QQ,f \rangle|\dvtx f\in\linf,{\|f\|
}_{\linf}\leq1\}$. We do not differentiate between the elements of
$\lone(\PP)$ and
their images under the natural bidual embedding $\lone\hookrightarrow
\ba$. In other words, we identify a countably additive measure $\QQ$
absolutely continuous with respect to $\PP$ with its Radon--Nikodym
derivative $\frac{d\QQ}{d\PP}$.

All of the spaces above admit natural positive cones, denoted by
$\lpee_+$, for $p\in[0,\infty]$ or $\ba_{+}$ in the case of
$\ba$. The domain of the pairing
$\langle\cdot,\cdot\rangle\dvtx\ba\times\linf\to\R$ can be
replaced by
$\ba_+\times\lzer_+$ by setting $\langle\QQ,f \rangle:=\lim
_{n\to\infty}
\langle\QQ,f\wedge n \rangle\in[0,\infty],\mbox{ for }\QQ\in\ba
_+\mbox
{ and }
f\in\lzer_+$. Each element $\QQ\in\ba_+$ admits the unique
decomposition (called the Yosida--Hewitt decomposition)
$\QQ=\QQ^r+\QQ^s$ into a countably-additive measure $\QQ^r\in\lone_+$
and a singular part $\QQ^s\in\ba_+$ uniquely characterized by the fact
that $\QQ'\equiv0$, whenever $\QQ'\in\lone_+$ and $\QQ'(A)\leq
\QQ^s(A)$ for all $A\in\FF$.

For an ordered normed space $N$ with the closed positive
orthant $N_+$ and \mbox{$y\geq0$}, we set $\bn(y):=\{x\in N\dvtx{\|x\|
}\leq y\}$, $\bn_+(y):=\bn(y)\cap N_+$, $\sn(y):=\{x\in N\dvtx{\|
x\|}= y\}$ and $\sn_+(y):=\sn(y)\cap N_+$. For a dual pair
$(X,X^*)$ of vector spaces (with the pairing denoted by
$\langle\cdot,\cdot\rangle$) and a map $f\dvtx X\to(-\infty,\infty
]$, $f^*$
denotes the $(X,X^*)$-convex conjugate of $f$, that is,
$f^*(y):=\sup_{x\in X} ( \langle x,y \rangle-f(x) )$, $y\in X^*$.
Finally, we remind the reader that the (convex-analytic) indicator
${\chi}_{B}$ of a subset $B$ of $X$ is defined by ${\chi}_{B}(x):= 0$ for
$x\in B$ and $+\infty$ otherwise.

\section{Problem formulation and the main results}
\subsection{The asset-price model}
We consider a financial market with $d\in\N$ risky assets modeled by a
$d$-dimensional c\`{a}dl\`{a}g semimartingale
\[
S=
\bigl(S^{(1)}_t,\ldots,S^{(d)}_t\bigr)_{t\in[0,T]}.
\]
The existence of a num\'eraire
asset $(S^{(0)}_t)_{t\in[0,T]}$, with $S^{(0)}_t:=1$ for
$t\in[0,T]$---a~zero-interest money-market account---is also postulated.

A predictable $S$-integrable process
$H=(H^{(1)}_t,\ldots,H^{(d)}_t)_{t\in[0,T]}$ is called a
\textit{portfolio} and its value $H_t$ is
interpreted as the number of shares of each risky asset held by the
investor at time $t\in[0,T]$. If a portfolio $H$ is used to implement a
dynamic trading strategy, the gains/losses accrued by time $t$ are
given by
$X^H_t$, where
%
\begin{equation}\label{defwealth}
X^{H}_t:= (H \cdot S)_t:= \int_0^t \sum_{k=1}^d
H^{(k)}_udS^{(k)}_u,\qquad t\in[0,T].
\end{equation}
The sum on the right has to be understood in the sense of
vector stochastic integration; see~\cite{CheShi02,Jac80} and
Chapter VII, Section 1a, in~\cite{Shi99}.

\subsection{Convex constraints}
\label{sseconv-const}
Let $\CR$ denote the set of all nonempty closed and convex subsets
of $\R^d$.
%
\begin{definition}
\label{defpredictable}
A map $\kappa\dvtx[0,T]\times\Omega\to\CR$ is said to be
\textit{predictable} if the set
\[
\{(t,\omega)\in[0,T]\times\Omega\dvtx\kappa(t,\omega)\cap F\ne
\varnothing\}
\]
is predictable for each closed set $F\subseteq\R^d$.
\end{definition}

We fix a predictable \textit{constraint map} $\kappa\dvtx[0,T]\times
\Omega\to\CR$; it is
used as a specification of an exogenously-imposed constraint on the
possible values the portfolio $H$ can take. The set of all portfolios
$H$ such that $H_t\in\kappa_t$ for all $t\in[0,T]$, $\PP$-a.s., will
be denoted by $\sA^\kappa$. The investment in the money market account
is not restricted.

In addition to the constraint imposed through $\kappa$, we consider a
different kind of a constraint known as the \textit{admissibility
constraint}. More precisely, a portfolio $H$ for which the process
$X^H:= H\cdot S$ there exists a constant $a\geq0$ such that
$X^H_t\geq-a$ for all $t\in[0,T]$, $\PP$-a.s., is called
\textit{admissible}. Such a constraint is commonplace in mathematical
finance and is imposed to rule out doubling strategies. The set of all
admissible portfolio processes is denoted by $\sA^{\mathrm{low}}$.

Combining the above the two constraints produces the class $\A$ of
\textit{constrained admissible portfolios},
\[
\A:=\sA^{\mathrm{low}}\cap\sA^\kappa.
\]
Many classical constraint
structures can be expressed in terms of a well-chosen $\kappa$; see,
for example, Section 3 in~\cite{Cuo97} and Chapter 5 in \cite
{KarShr98}. We do exhibit, however, in some
detail the construction that allows us to treat the presence of a
random endowment in our framework:
%
\begin{example}[(Random endowment as a special case of a portfolio
constraint)]\label{exranendow}
As above, let the financial market consist of the risky assets $S=
(S^{(1)}_t,\ldots,S^{(d)}_t)_{t\in[0,T]}$ and the riskless asset
$S^{(0)}_t:= 1$. Let us also assume that $S$ admits no arbitrage
in the sense of the condition NFLVR. Consequently, there exists an
equivalent $\sigma$-martingale measure $\QQ$; see~\cite{DelSch98}
for the terminology.

Let us also assume that the agent receives a lump-sum \textit{random
endowment} $\EN\in\linf(\FF_T)$ at time $T$. For an arbitrary
equivalent $\sigma$-martingale measure $\QQ$, the process
$\hat{S_t}$, defined as a c\`{a}dl\`{a}g version of the bounded martingale
$\EE^{\QQ}[\EN|\FF_t]$, can be
added to $S$ to form a larger financial market. The constraint set
$\kappa$ is defined so as to mimic the behavior in the original
market with the presence of the random endowment,
\[
\kappa_t(\omega):= \R^d\times\{1\}.
\]
Indeed, any admissible constrained portfolio in the augmented market
$(S,\hat{S})$ leads to a total wealth of the form $(H\cdot
S)_T+\hat{S}_T-\hat{S}_0= x+(H\cdot S)_T + \EN$, for
$x:=-\EE^{\QQ}[\EN]$ (under the assumption that $\FF_0$ is
$\PP$-trivial). Thanks to the boundedness of~$\EN$, the notions of
admissibility in the two markets are equivalent.

It is possible to extend the domain of this example in various
directions. For example, to treat an unbounded random endowment,
one would need to use a more sophisticated
version of the admissibility requirement or resort to a change of
num\'eraire.
\end{example}

\subsection{No-arbitrage conditions on the financial market} Moving
on toward our main result, we introduce notation for the set of
gains processes of admissible constrained portfolios, as well as for
certain related sets,
%
\begin{eqnarray}
\label{defC}
\Xc&:=&\bigl\{(X^H_t)_{t\in{[0,T]}}\dvtx H\in\A\bigr\},\nonumber\\
\sK&:=&\{X_T\dvtx X\in\Xc\},\nonumber\\[-8pt]\\[-8pt]
\sC&:=&(\sK-\lzer_+)\cap\linf,\nonumber\\
\sC(x)&:=&(x+\sK-\lzer_+)\cap\lzer_+\qquad\mbox{for }x\in\R.\nonumber
\end{eqnarray}
As far as technical conditions are concerned, we start with a
succinct umbrella assumption under which our main
theorem holds. Natural sufficient conditions on separate
ingredients---the market and the constraint correspond\-ence---will be briefly
described below, and then in detail in Section
\ref{secsuff}.

Following~\cite{Zit09}, we say that a subset of a topological vector
space is \textit{convexly compact} if any family of closed and
convex sets with the finite-intersection property admits a nonempty
intersection. In~\cite{Zit09}, it is shown that a subset of
$\lzer_+$ is convexly compact if and only if it is bounded and
closed in probability.
%
\begin{assumption}\label{assC}
$\CC(x)$ is convexly compact for all $x\in\R$, and there exists
$x\in\R$ such that $\CC(x)\ne\varnothing$.\vadjust{\goodbreak}
\end{assumption}
%
\begin{remark}
Let us comment on the interpretation of Assumption
\ref{assC}. The nonemptiness condition is equivalent to assuming
$\sA\neq\varnothing$, that is, that it is possible to produce a
bounded-from-below wealth process without violating the
constraints. Boundedness in probability serves as a weak
no-arbitrage requirement and can be deduced, in may cases, already
from the finiteness of the expected-utility value function. Similar
weakenings of the no-arbitrage condition have already been
considered in the literature; see, for example, Section 3 in
\cite{KarKar07}. The closedness requirement is a natural
condition for the existence of an expected-utility optimizer and is
present in virtually all widely-used no-arbitrage concepts.
\end{remark}

Let us preview a sufficient condition for Assumption
\ref{assC}. The definitions of the map $\Pi^S$ (the
\textit{projection onto
the predictable range map} of~\cite{CziSch11}) and the support measure
$\PP^S$ of $S$ are postponed until Section~\ref{thmclosed}. Let us
mention that the below condition (2) is always
satisfied for any $S$ if $\kappa_t(\omega)$ is polyhedral, compact,
or if
it admits a continuous support function, for each $t\in[0,T]$,
$\PP$-a.s.; see~\cite{CziWesZhe11} for details.
However,~\cite{CziSch11} and~\cite{CziWesZhe11} contain examples
showing that (2) in the following proposition is not true in general.

%
\begin{proposition}
\label{prosufficient}
Assumption~\ref{assC} holds if the following three conditions are satisfied:
\begin{enumerate}
\item[(1)] $\sA\ne\varnothing$;
\item[(2)] the projection $\Pi_t^S(\omega)\kappa_t(\omega)$ is
closed, for
$\PP^S$-a.e.;
\item[(3)] there exist:
\begin{enumerate}[(a)]
\item[(a)] a probability measure $\QQ\sim\PP$;\vspace*{1pt}

\item[(b)] $\hH\in\AA$ with
$\EE^{\QQ}[ (\hH\cdot S)_T]<\infty$ and $\hH\cdot S$ locally bounded;
\item[(c)] a nondecreasing predictable c\`{a}dl\`{a}g process $\{ A_t \}
_{t\in[0,T]}$, with $A_0=0$,
\end{enumerate}
such that
%
\begin{equation}\label{equsup-cond}
H\cdot S - (\hH\cdot S + A) \mbox{ is a
$\QQ$-supermartingale}\qquad \mbox{for all $H\in\AA$.}
\end{equation}
\end{enumerate}
%
\end{proposition}
%
\begin{remark}
\label{remsuff-pro-suff}
(1) Conditions on the constraint set $\kappa$, under which
property~(2) in
Proposition~\ref{prosufficient} holds, are presented in~\cite{CziSch11}.

(2) The process $A$ in (3)(c) above is allowed to depend on the measure
$\QQ$ from (3)(a) and the process $\hH$ from (3)(b). It has to guarantee
the supermartingale property of $H\cdot S- (\hH\cdot S+A)$, however, for
all $H\in\AA$ simultaneously.

(3) In the unconstrained case, the existence of a local-martingale
measure for $S$ suffices for property (3) in the above proposition with
$A=0$ and
$\hH=0$. When the constraint set forms a convex cone, the process $A$
scales away (unlike in~\cite{KarKar07} where the admissibility
criterion is
different), and the existence of a local supermartingale measure suffices.

(4) The supermartingale requirement in Proposition~\ref{prosufficient}
(3)(c) can be weakened by imposing additional regularity
on $A$ and $\hH\cdot S$. More precisely, if $A_T$ is $\QQ$-integrable
and $\hH\cdot S$ is a $\QQ$-uniformly integrable martingale,
it is enough to assume that the process $H\cdot S-(\hH\cdot S+A)$
is a $\QQ$-\textit{local} supermartingale. Indeed, the (full)
$\QQ$-supermartingality will then immediately follow by the (DL) property
of its negative part.
\end{remark}

We conclude this section with an example in a ``Brownian'' setting.
%
\begin{example}[(It\^{o}-process-driven models)]
Let us consider the standard It\^{o}-process setting used, for example, in
\cite{KarShr98}. We fix $d\in\N$ and let $(W_t)_{t\in[0,T]}$ be a
$d$-dimensional Brownian motion and $(\sF_t)_{t\in[0,T]}$ its augmented
filtration. The stock price dynamics are given by
%
\begin{equation}
\label{equito-S}
dS_t:=
\mu_t\,dt + \sigma_t \,dW_t,\qquad S_0:=1,\qquad t\in[0,T],
\end{equation}
with the
$d$-dimensional column-vector process $(\mu_t)_{t\in[0,T]}$ and the
$d\times d$-matrix process $(\sigma_t)_{t\in[0,T]}$ are progressively
measurable, and such that the integrals in (\ref{equito-S}) are well defined.

With no invertibility requirements imposed
on it, $\sigma_t$ can be assumed to be a square matrix,
that is, that there are as many risky assets as there are independent
Brownian motions, without loss of generality. For later use, we define
the linear-subspace-valued
process $(I_t)_{t\in[0,T]}$---called the \textit{span process}---by
\[
I_t:= \{\sigma_t \nu\dvtx\nu\in\R^d\}.
\]

As far as the constraints are concerned, we fix a closed convex constraint
map $(\kappa_t)_{t\in[0,T]}$ and associate to it is the
\textit{recession-cone process} $(R_t)_{t\in[0,T]}$ defined by
\[
R_t:=\{\xi\in\R^d\dvtx\forall t>0, \exists y\in\kappa_t,
y+t\xi\in\kappa_t\}.
\]
In words, $R_t$ contains all the directions in
which $\kappa_t$ is unbounded. We will also need the \textit{barrier-cone
process} whose values are the polar cones of the values of~$R_t$, that is,
\[
B_t:=\{\eta\in\R^d\dvtx\eta^T \xi\leq0\mbox{, for all }\xi\in
R_t\}.
\]
Consider now the following condition:
%
\begin{equation} \label{equcond1}
I_t \cap( \mu_t - B_t ) \ne\varnothing\qquad
\mbox{on } \Omega\times[0,T].
\end{equation}
In words, either $\mu_t$ is contained in the image of $\sigma_t$ (the
typical no-arbitrage requirement in the unconstrained case), or we can
travel to $\mu_t$ from some point in the image of $\sigma_t$ using one
of the elements of the barrier cone as a velocity vector. We note that,\vadjust{\goodbreak}
by choosing an appropriate constraint structure, one can, without loss
of generality, assume that $(\sigma_t)_{t\in[0,T]}$ is\vspace*{1pt}
everywhere invertible, and, thus, that $I_t=\R^d$. For flexibility's
sake, we opt to keep both processes at the current level of generality.

The correspondence $(t,\omega)\to I_t(\omega) \cap
(\mu_t(\omega)-B_t(\omega)))$ takes values in the set of
nonempty closed subsets of $\R^d$. Moreover, it is
weakly measurable with
respect to the progressive $\sigma$-algebra; see Definition
18.1, page 592, of~\cite{AliBor06} for various measurability notions for
correspondences. Indeed, this follows easily from
the progressive measurability of the processes $\mu$ and $\sigma$.
Therefore, we can apply the Kuratowski--Ryll--Nardzewski Selection
theorem (see Theorem 18.13, page~600, in~\cite{AliBor06}) which guarantees
the existence of a progressively measurable process
$(\hat{\mu}_t)_{t\in[0,T]}$ with $\hat{\mu}_t\in I_t \cap(\mu_t -
B_t)$. Then, we can pick a process $(\nu_t)_{t\in[0,T]}$ such
that $\sigma_t \nu_t = \hat{\mu}_t$. This can be done, for example,
through the
(measurable) operation of choosing the unique minimal-norm solution of
a solvable linear system, that is, by taking the Moore--Penrose
inverse; see page 35 of~\cite{BenGre03} for definitions and example 25
on page 101 for
the statement and the proof of the so-called Tihonov-regularization
representation which can be used to deduce the aforementioned
measurability of the Moore--Penrose pseudoinversion.

Assuming\vspace*{1pt} that the stochastic exponential $\EN(-\nu\cdot
W)$ is a (true) martingale, we define the measure $\QQ^{\nu}\sim\PP$ by
$\frac{d\QQ^{\nu}}{d\PP}:=\EN(-\nu\cdot W)_T$. For two processes
$H$, \mbox{$\hH\in \AA$}, we note that finite-variation part in the
semimartingale decomposition of the process $(H-H')\cdot S$ under the
probability measure $\QQ^{\nu}$ is absolutely continuous with the
derivative given by
\[
(H_t-H'_t)^T(\mu_t - \sigma_t \nu_t)=
(H_t-H'_t)^T\beta_t.
\]
Since $\beta_t\in B_t$, one can find the
``farthest'' point in $\kappa_t$ in the direction $\beta_t$. More
precisely, we set $\hH_t=\argmax_{h\in\kappa_t} h^T \beta_t$.
Then,\vspace*{-1pt}
it follows that $(h-\hH_t)(\mu_t-\sigma_t \nu_t)\leq0$, for all
$h\in
\kappa_t$. If one could ensure that the so-constructed process
$(\hH_t)_{t\in[0,T]}$ indeed belongs to the admissible set
$\AA^{\mathrm{low}}$ and that $\hH\cdot S$ is a $\QQ^{\nu}$-martingale,
part (4) of Remark~\ref{remsuff-pro-suff} would guarantee
that the requirement (3) in
Proposition~\ref{prosufficient} is fulfilled in a very parsimonious
way: we
could simply take $A_t:= 0$.

Alternatively, one can exchange some of the unpleasant regularity
needed for the
above approach for the necessity of the use of a nontrivial process $A$.
Indeed, let the processes $\nu$ and $\beta$ be as above, and let
$\hH\in\AA$ be such that $\hH\cdot S$ is a $\QQ^{\nu}$-martingale;
$\hH_t:= 0$ is always a possibility.

We define the process $A$ as
\[
A_t:= \int_0^t
\bigl( \delta_{\kappa_u}(\beta_u)- \hH_u^T \beta_u \bigr) \,dt,
\]
where $\delta_{\kappa_t}(\xi):=\sup_{h\in\kappa_t} h^T \xi$ is
the support
function of the constraint set $\kappa_t$. This way,
we can fulfill
requirement (3) in Proposition~\ref{prosufficient}, by checking that
$\EE^{\QQ^{\nu}}[ A_T]<\infty$.

Finally, let us shortly describe a case in which no equivalent
local-martin\-gale measure can be found in the unconstrained version of the
market, but one can still verify the conditions of Proposition
\ref{prosufficient}. We take $d:=1$, $\sigma_t:= 1$ and
a progressively-measurable process $\mu_t$ such that:
\begin{longlist}[(2)]
\item[(1)] $\EE[\int_0^T \mu_t^2\,dt]<\infty$, but
\item[(2)] $\EE[ \EN(-\mu\cdot
W)_T]<1$; that is,
$\EN(-\mu\cdot W)$ is not a true martingale.
\end{longlist}
An example of such a process $\mu_t$ can be based on the
three-dimensional Bessel
process; see, for example, Example 2.2 in~\cite{Lar09} for details.
Girsanov's theorem implies that no local-martingale measure can exist for
$S$.
Indeed, the only candidate fails to be a probability measure.

On the other hand, let us choose a constant constraint set $\kappa_t:=
[-1,1]$ and take $\QQ:=\PP$, $\hH:= 0$ and $ A_t:= \int_0^t
\delta_{\kappa_t}(\mu_t) \,dt = \int_0^t |\mu_u| \,du$. For any
$H\in\AA$, we have
\[
(H\cdot S)_t - A_t = \int_0^t (H_u \mu_u -
|\mu_u| ) \,du + \int_0^t H_u \,dW_u,
\]
a process which is clearly
a supermartingale. Consequently, the conditions of proposition (3) are
satisfied.

A more extreme version of the above can be constructed by simply taking
$S_t:=t$ and $\kappa_t:= (-\infty,1]$. The original, unconstrained,
market allows for
(unbounded) arbitrage which cannot be implemented without violating the
constraints. Constraints still allow for a limited riskless gain, but
the conditions of Proposition~\ref{prosufficient}(3) hold.
\end{example}



\subsection{The primal problem}
The investor's preferences are modeled by a function $U$---called a
\textit{utility function}---which will always be assumed to satisfy the
following
assumption:
%
\begin{assumption}
\label{assU}
$U\dvtx(0,\infty)\to\R$ is a nondecreasing and concave function with the
following two properties:
%
\begin{eqnarray}%
\label{equRAE}
&&\exists x_0>0, c\in(1,2)\ \forall x \geq x_0\nonumber\\
&&\qquad
U(2x)\leq c U(x),\qquad
\lim_{x\searrow0} U'^{+}(x)=\infty\hspace*{40pt}\\
&&\eqntext{\mbox{where $U'^{+}$ denotes the
right derivative.}}
\end{eqnarray}
\end{assumption}
%
\begin{remark}
The first part of condition (\ref{equRAE}) is a derivative-free
restatement of the
notion of the reasonable asymptotic elasticity of~\cite{KraSch99} (for
details, see Lemma 6.3(i) in~\cite{KraSch99}), and it restricts the rate
of growth of $U$ in the neighborhood of $+\infty$. In
particular, (\ref{equRAE}) implies that the Inada condition at
$+\infty
$, namely, $\lim_{x\to\infty}
U'(x)=0$, is satisfied if $U'$ is interpreted as either the left or the right
derivative.
\end{remark}

To simplify the notation later on, we extend the definition of $U$ by
semicontinuity to $[0,\infty)$ by setting $U(0)= \inf_{\xi>0} U(\xi)$
and, further, to $\R$, by $U(x)=-\infty$, for $x<0$. The
(\textit{primal}) \textit{value function} $u\dvtx\R\to[-\infty,\infty
]$ of the
utility-maximization problem, parametrized by the investor's initial
wealth $x\in\R$, is then defined by
%
\begin{equation}\label{defu}
u(x):= \sup_{X\in\sK} \EE[U(x+X)],
\end{equation}
where we use the convention that for $\xi\in\lzer$, one has $\EE
[\xi]
= -\infty$ whenever $\EE[\xi^-]=\infty$, even if $\EE[\xi
^+]=\infty$.

The monotonicity of $U$ and the fact that $U(x)=-\infty$ for
$x<0$, imply
\[
u(x)=\sup_{X\in(\sK-\lzer_+)} \EE[
U(x+X)]=\sup_{f\in\sC(x)} \EE[ U(f)],
\]
where
$\sup\varnothing:=-\infty$. The monotone convergence theorem guarantees
that
\[
u(x)=\sup_{f\in\sC} \UU(x+f),
\]
where
the map $\UU\dvtx\linf\to[-\infty,\infty)$ is
a shorthand for $f\mapsto\EE[U(f)]$ with $U$ regarded as defined on
$(-\infty,\infty)$.

\subsection{The dual problem}
To introduce the dual optimization problem we first need to recall the
notion of a support function. Let $\sC$ be as in (\ref{defC}) above,
and let $\sP$ denote the set of all (countably additive) probability
measures on
$(\Omega,\FF)$ which are absolutely continuous with respect to $\PP$.
The \textit{support function $\ac$ of $\sC$} is defined by
%
\begin{equation}
\label{equsupport}
%
\sP\ni\QQ\to\ac(\QQ):=\sup_{f\in\sC} \EE^{\QQ}[f]\in
(-\infty
,\infty].
\end{equation}

The optimization problem (for now only formally) dual to the primal
utility-maximization problem (\ref{defu}) above is defined by
its value function $v\dvtx[0,\infty)\to[-\infty,\infty]$,
%
\begin{equation}\label{defdual}
v(y):= \inf_{\QQ\in\sP}\biggl( \EE\biggl[ V\biggl(y \,\frac{d\QQ}{d\PP}\biggr)\biggr] +
y\ac(\QQ)\biggr),
\end{equation}
where
$V(y):= \sup_{x\in\R} ( U(x) - xy )\in
(-\infty,\infty]$, $y\in\R$, is the Fenchel--Legendre
transform of $-U(-\cdot)$.

\subsection{Main result}
The following theorem extends some of the main existence results in
\cite{CviSchWan01,KraSch99} and~\cite{KraSch03} to the
constrained case and shows that countably-additive measures suffice to
describe the dual value function.
%
\begin{theorem}\label{thmmain} Let $u$ and $v$ be defined by
(\ref{defu}) and (\ref{defdual}), respectively, and assume
that $u(x)\in\R$ for some $x\in\R$. Under
Assumptions~\ref{assC} and~\ref{assU},
with $\underline{x}:=\inf\{x\in\R\dvtx u(x)>-\infty\}$,
the following assertions
hold:
\begin{longlist}[(4)]
\item[(1)] The function $u$ is concave, upper semicontinuous and
nondecreasing, while $v$ is convex and lower semicontinuous.\vadjust{\goodbreak}
\item[(2)] We have $\ux=-\inf_{\QQ\in\sP} \ac(\QQ)$, where $\ac
$ is
defined in
(\ref{equsupport}). Furthermore, $u(x)\in\R$ for
$x\in(\ux,\infty)$ and $u(x)=-\infty$ for $x\in(-\infty,\ux)$.
\item[(3)] For each $x\in\R$ with $u(x)\in\R$ (and, in particular,
for $x>\underline{x}$), there exists $\Hx\in\A$ such that
\[
u(x)=\EE\biggl[ U\biggl(x+\int_0^T \Hx_u \,dS_u\biggr)\biggr].
\]
\item[(4)] The following conjugacy relations hold:
%
\begin{eqnarray}
\label{equconjugate}
v(y) &=& \sup_{x\in\R} \bigl( u(x) - xy\bigr),\qquad y\in\R,\\
\label{eqvconjugate}
u(x) &=& \inf_{y\in[0,\infty)} \bigl( v(y) + xy\bigr),\qquad
x\in\R.
\end{eqnarray}
\end{longlist}
\end{theorem}
%
\begin{remark}
(1) Theorem
\ref{thmmain} and Example~\ref{exranendow} show that Theorem~2.2(iv) in
\cite{KraSch99} indeed carries over to the random-endowment setting of~\cite{CviSchWan01}
also when the utility function $U$ is nonsmooth.
Theorem 3.1(ii) in~\cite{CviSchWan01} provides a link between the
primal and dual optimizers. As we discuss in the next section, we can
only guarantee the existence of a finitely additive dual minimizer
$\hat{\QQ}_y\in\ba$ and, in general, we will not have
$\hat{\QQ}_y\in\sP$. Under the additional assumption that $U$ is
strictly concave, the dual function $V$ is differentiable by
Theorem 26.3 in~\cite{Roc70}. We can then extend Theorem 3.1(ii) in
\cite{CviSchWan01} to our setting by using the Yosida--Hewitt
decomposition of $\QQ\in\ba$ into its regular part $\QQ^r\in\lone_+$
and its purely singular part $\QQ^s$ as follows. For
$x>\sup_{\QQ\in\sP} -\ac(\QQ)$ we have the relation
%
\begin{equation}\label{eqprimaldual} x + \int_0^T \Hx_u \,dS_u = -
V'\biggl(\hat{y} \,\frac{d\hat{\QQ}_{\hat{y}}^r}{d\PP}\biggr),
\qquad \PP\mbox{-a.s.,}
\end{equation}
where
$\hat{y}$ attains the infimum in (\ref{equconjugate}), and
$\hat{\QQ}_{\hat{y}}^r$ denotes the regular part of $\hat{\QQ
}_{\hat{y}}$,
a minimizer in the generalized dual problem $v^{\mathrm{ba}}$; see Section
\ref{sec3.1} for details. By using the positive homogeneity of the support
function $\alpha_\sC$, the proof of (\ref{eqprimaldual}) is a
straightforward application of the ideas in~\cite{CviSchWan01}.

(2) When $U$ is not necessarily strictly concave,
\cite{BouTouZeg04} and later~\cite{WesZhe09} establish the validity of
(\ref{eqprimaldual}) in the setting of Example~\ref{exranendow} when
$V'$ is replaced by the $V$'s subdifferential $\partial V$. However, as
discussed in their Remark 3.9.3, the authors of~\cite{BouTouZeg04}
assume a specific relationship between the domain of $U$ and the norm
$\|\EN\|_{\linf(\sF_T)}$, which makes it difficult to compare their
setting to ours. Finally, we mention Westray and Zheng~\cite{WesZhe11}
who illustrate a possible pitfall related to using $\partial V$ instead
of $V'$ in (\ref{eqprimaldual}) when $U$ is not strictly
concave.\looseness=-1
\end{remark}

\section{Proofs}


\subsection{A relaxation of the dual problem}\label{sec3.1}
We first note that $\ac$ naturally extends from $\sP$ to the space
$\ba$ by replacing the expectation $\EE^{\QQ}[f]$ by the value\vadjust{\goodbreak}
$\langle\QQ,f \rangle$ of the dual pairing in (\ref{equsupport}).
With such
an extended domain, $\ac$ coincides with the convex
$(\linf,\ba)$-conjugate $({\chi}_{\sC})^*$ of the convex indicator
${\chi}_{\sC}$. It
follows, in particular, that $\ac$ is convex and $\sigma(\ba,\linf)$-lower
semicontinuous.

To extend the dual value function, we
follow~\cite{Zit05} and define the map
$\VV\dvtx\ba\to(-\infty,\infty]$ of $\UU$ by
%
\begin{equation}
\label{genV}
%
\VV(\QQ):= \sup_{f\in\linf}\bigl(
\UU(f) - \langle\QQ,f \rangle\bigr)\qquad \mbox{for }\QQ\in\ba.
\end{equation}
We note that $\VV=\hUU^*$, for the $(\linf,\ba)$-duality, where
$\hUU(f)=-\UU(-f)$. A~minimal modification of Lemma 2.1, page 138, in
\cite{OweZit09} produces the following representation:
%
\begin{equation}
\label{equV-reg}
\VV(\QQ)=
\cases{\displaystyle
\EE\biggl[V\biggl(\frac{d\QQ^r}{d\PP}\biggr)\biggr],&\quad
$\QQ\in\ba_+$,\vspace*{2pt}\cr \infty, &\quad
$\QQ\notin\ba_+$.}
\end{equation}
As mentioned in the \hyperref[intro]{Introduction}, $\QQ^r$ denotes the regular
part in the Yosida--Hewitt decomposition $\QQ= \QQ^r + \QQ^s$.

With $\VV$ and $\ac$ extended as above, a relaxed version of
the dual value function can be posed over the $y$-sphere in $\ba$:
\[
\vba(y):= \inf_{\QQ\in\sbayp}\bigl( \VV(\QQ) + \ac(\QQ)
\bigr)\qquad
\mbox{for }y\geq0.
\]
Since $y\sP$ can be identified with $\sl_+(y)$,
which, in turn, admits a natural embedding into $\sbayp$, it is clear
that $\vba(y)\leq v(y)$. It is the equality between the two functions
(as demonstrated in Proposition~\ref{prov-vba} below) that will be
one of the major steps in the proof of our main Theorem
\ref{thmmain}. Unfortunately, it is not true in general that the
involved quantities are $\sigma(\ba,\linf)$-upper semicontinuous so
this equality
cannot be deduced from the $\sigma(\ba,\linf)$-density of $\sl(y)$
in $\sba(y)$.

Some of the advantages that working with $\vba$ affords over $v$ are
evident from the
following result, which follows directly from the $\sigma(\ba,\linf
)$-compactness
of $\sbayp$ (the Banach--Alaoglu theorem) and the $\sigma(\ba,\linf)$-lower
semicontinuity of $\VV+\ac$.
%
\begin{proposition}
\label{profirst-uv}
If $\sC\ne\varnothing$ and Assumption~\ref{assU} holds,
$\vba(y)$ admits a minimizer for each $y> 0$. More precisely,
there exists $\hQy\in\sbayp$ such that
$\vba(y)=\VV(\hQy)+\ac(\hQy)$.
\end{proposition}

\subsection{Conjugacy of value functions}
%
\begin{proposition}
\label{propconjugates}
Suppose that Assumptions~\ref{assC} and~\ref{assU} hold and that
$u(x) \in\R$ for some
$x\in\R$. Then:
\begin{longlist}[(2)]
\item[(1)] $\vba(y)=\sup_{x\in\R} ( u(x)-xy )$, for all
$y\in\R$,
and
\item[(2)] there exists $y>0$ such
that $\vba(y)<\infty$.\vadjust{\goodbreak}
\end{longlist}
\end{proposition}
\begin{pf}
(1) By the Banach--Alaoglu theorem, $\sbayp$ is $\sigma(\ba
,\linf)$-com\-pact for
any $y\geq0$. Moreover, the Lagrangian,
\[
L(\QQ,(f,g)):=\UU(f)-\langle\QQ,f-g \rangle\mbox{:}
\]

\begin{longlist}[(a)]
\item[(a)] is concave in $(f,g)$ on $\linf\times\linf$, and
\item[(b)] convex, and $\sigma(\ba,\linf)$-lower semicontinuous in $\QQ$ on
$\ba$.
\end{longlist}
Therefore, the minimax theorem (see~\cite{Sio58}) can be used
to interchange $\inf$ and $\sup$ in (\ref{equchain}) below. Also,
let us note that for $h\in\linf$ and $y\geq0$, we have
\[
\sup_{\QQ\in\sbayp} \langle\QQ,h \rangle=y\esssup h.
\]
%
It follows that
%
\begin{eqnarray}%
\label{equchain}
\vba(y) &=&\inf_{\QQ\in\sbayp} \Bigl( \VV(\QQ)+\sup_{g\in\sC}
\langle\QQ,g \rangle\Bigr)\nonumber\\
&=&
\inf_{\QQ\in\sbayp
}\sup_{f\in
\linf}
\Bigl( \UU(f)-\langle\QQ,f \rangle+\sup_{g\in\sC} \langle\QQ
,g \rangle\Bigr)
\nonumber
\\
&=&\inf_{\QQ\in\sbayp}\sup_{(f,g)\in\linf\times\sC} \bigl(
\UU(f)-\langle\QQ,f-g \rangle\bigr)\\
& =&\sup_{(f,g)\in
\linf
\times\sC}
\inf_{\QQ\in\sbayp} \bigl( \UU(f)-\langle\QQ,f-g \rangle\bigr)
\nonumber\\
&=&
\sup_{(f,g)\in\linf\times\sC} \bigl( \UU(f)-y
\esssup(f-g)\bigr).\nonumber
\end{eqnarray}
We can split the last supremum according to the value of $\esssup
(f-g)$ and use the monotonicity of $\UU$ to obtain
\begin{eqnarray*}
\vba(y)&=&\sup_{x\in\R} \sup_{g\in\sC, f\in\linf f\leq g+x}
\bigl(\UU(f)-yx \bigr)\\
&=&\sup_{x\in\R} \sup_{g\in\sC} \bigl(
\UU(x+g)-yx\bigr)=\sup_{x\in\R} \bigl( u(x)-xy \bigr).
\end{eqnarray*}

(2) This is a direct consequence of the standing assumption
that $u$
is proper and the fact that properness is preserved under conjugacy; see
Theorem 12.2, page 104, in~\cite{Roc70}.
\end{pf}

\subsection{Existence in the primal problem}
We start with a variant of the argument developed in the proof of
Theorem 4.2 in~\cite{DelSch94}, adjusted to our case of convex
constraints.
%
\begin{lemma}
Under Assumption~\ref{assC},
the set $\sC$ is nonempty,
and $\sigma(\linf,\break\lone)$-closed.
\end{lemma}
\begin{pf}
Let $x\in\R$ be such that $\sC(x)$ is nonempty. Then, there
exists $X\in\sK$ such that $x+X\geq0$, $\PP$-a.s., and so the
constant random
variable $-x$ belongs to $\sC$, proving that $\sC$ is nonempty.\vadjust{\goodbreak}

To prove closedness, for $M>0$ we define the closed $\linf$-ball
$B^{\linf}(M)= \{f\in\linf\dvtx \|f\|_\infty\le M\}$. By a version of
Grothendieck's lemma (see, e.g., Theorem 5.1 in~\cite{FolKra97}) and
the convexity of $\sC$, the claim is equivalent to showing\vspace*{1pt}
that $\sC\cap B^{\linf}(M)$ is closed in probability for all $M>0$. So
let $(f_n)_{n\in\N} \subset \sC\cap B^{\linf}(M)$ converge to $f_0$ in
probability. It is clear that $f_0 \in B^{\linf}(M)$, and we only need
to show that $f_0\in\sC $. We have $f_n + M\ge0$, hence, $f_n
+M\in\sC(M)$. By Assumption \ref {assC}, the set $\sC(M)$ is closed in
probability which ensures that $f_0+M\in\sC(M)$; that is, there exists
$H\in\sA$ such that $f_0 +M\le M + (H \cdot S)_T$. Therefore,
$f_0\in\sC$.\vspace*{-2pt}
\end{pf}

By using the extended definition $\langle\QQ,f \rangle:= \lim
_{n\to\infty}
\langle\QQ,f\land n \rangle$ for $\QQ\in\sP$ and $f\in\mathbb{L}^0_+$,
we have
the following characterization of the sets $\sC$ and $\sC(x)$.\vspace*{-2pt}

\begin{corollary}
\label{corhedge}
Under Assumption~\ref{assC}:
\begin{longlist}[(2)]
\item[(1)]
$f\in\linf$ belongs to $\sC$
if and only if $\langle\QQ,f \rangle\leq\ac(\QQ)$, for all $\QQ
\in\sP$;

\item[(2)] $f\in\mathbb{L}^0_+$ belongs to $\sC(x)$
if and only if $\langle\QQ,f \rangle\leq x+\ac(\QQ)$, for all $\QQ
\in\sP$.\vspace*{-2pt}
\end{longlist}
\end{corollary}
\begin{pf}
Closedness of $\sC$ implies that the convex function $\chi_{\sC}$ is
lower semicontinuous for the
$\sigma(\linf,\lone)$-topology. Therefore, $\chi_{\sC}$ is its own
$\sigma(\linf$, $\lone)$-biconjugate, and consequently,
$\chi_{\sC}(f)=\sup_{\QQ\in\sP}( \langle\QQ,f \rangle
-\ac(\QQ
))$
which proves~(1).

\mbox{For (2) we pick $f\!\in\!\sC(x)$ and $n\!\in\!\N$, and note that for some
$H\!\in\!\sA$, we have}
\[
(f-x)\land
n\le(H\cdot S)_T\land n\in\sC.
\]
Therefore, for $\QQ\in\sP$, Fatou's
lemma implies that
\[
\langle\QQ,f-x \rangle\le\liminf_{n\to\infty}
\langle\QQ,(f-x)\land n \rangle\le\ac(\QQ).
\]

Conversely, let $f\in\mathbb{L}^0_+$ be such that
$\langle\QQ,f-x \rangle\le\ac(\QQ)$ for all $\QQ\in\sP$. Then
for $n\in
\N$ we also
have $\langle\QQ,(f-x)\land n \rangle\le\ac(\QQ)$. Hence, by (1),
$(f-x)\land
n\in\sC$,
and so $(f-x)\land n +
x\in\sC(x)$, for all $n\in\N$. The claim now follows directly from the
closedness of $\sC(x)$
in probability.\vspace*{-2pt}
\end{pf}
%
\begin{lemma}
\label{lembd-tU}
$\!\!\!$Under Assumptions~\ref{assC} and~\ref{assU},
we have
$\sup_{f\in\sC(x)} \EE[U^+(f)]<\infty$, whenever
$u(x)\in\R$.\vspace*{-2pt}
\end{lemma}
\begin{pf}
We define the constant $x'=\inf\{x>0\dvtx U(x) \ge0\}$. If $x' = \infty$ there
is nothing to prove, and so, in what follows, we assume that
$x'\in[0,\infty)$. By Proposition~\ref{profirst-uv}, part (2), there
exist $y>0$ and
$\QQ\in\sbayp$ such that $\VV(\QQ)<\infty$ and $\ac(\QQ)<\infty$.
Since $\UU(f)\leq\VV(\QQ)+\langle\QQ,f \rangle$ for each
$f\in\linf$, in particular, for $f\in(x+\sC)\cap\linf_+$, we have
\begin{eqnarray*}
\EE[ U^+(f)]&\le&\EE\bigl[U\bigl(f {\mathbf1}_{\{f\geq x'\}}+x'{\mathbf1}_{\{
f<x'\}}\bigr)\bigr]\\
&\leq&
\VV(\QQ)+\langle\QQ,f \rangle+
\langle\QQ,x' \rangle\\
&\leq&
\VV(\QQ)+\ac(\QQ)+x'y,
\end{eqnarray*}
which is finite and independent of the choice of $f$.\vadjust{\goodbreak}
\end{pf}

Let us choose and fix constants $x_0>0$ and $c \in(1,2)$ as in Assumption~\ref{assU}.
For $h\in\linf$ with $h\ge x_0$, we then have
$U(2h)\leq c U(h)$; iterating this inequality produces
%
\begin{equation}
\label{equaaa} \EE[ U(2^mh)] \leq c^m\EE[ U(h)]\qquad
\mbox{for all } m\in\N, h\in x_0+\linf_+.
\end{equation}

\begin{proposition}
\label{propprimalexistence}
Under Assumptions~\ref{assC} and~\ref{assU},
for each $x\in\R$ with $u(x)\in\R$
there exists $\fx\in\sC(x)$ such that $u(x)=\UU(\fx)$.
\end{proposition}
\begin{pf}
The function $u$ is clearly concave, so the existence of $x\in\R$
such that
$u(x)<\infty$ implies that it is proper, that is, that $u(x)<\infty$,
for all
$x$. We pick $x\in\R$ with $u(x)<\infty$ and let
$\{f_n\}_{n\in\N}\subset\sC(x)$ be a maximizing sequence, that is,
a sequence in
$\sC(x)$ such that $\UU(f_n)\to u(x)$. Since $\sC(x)$ is bounded in
probability, we may find a sequence $\{g_n\}_{n\in\N}$, of convex combinations
$g_n\in\conv(f_n,f_{n+1},\ldots)$, which converges in probability to some
$\fx\in\lzer_+$. The concavity of $\UU$ implies that $g_n$ is also
a maximizing sequence. Furthermore, $\fx\in\sC(x)$ since $\sC(x)$
is closed
in probability.

To show that $\fx$ is indeed a maximizer, we use
Fatou's lemma to conclude that $\EE[ - U^-(\fx) ] \geq\limsup_n
\EE[ - U^-(g_n)]$, so that it is enough to show that $\EE[
U^+(g_n)]\to\EE[ U^+(\fx)]$. This will follow once we show that the
sequence
$\sq{U^+(g_n)}$ is uniformly integrable.

We start by defining the nonnegative constant
\[
x':= \inf\{ x>x_0\dvtx U(x) >0\}.
\]
If $x'=\infty$ there is nothing to prove, and so we
assume that $x'\in[0,\infty)$. We argue by contradiction and assume that
$\sq{U^+(g_n)}$ is not uniformly integrable. Lemma~\ref{lembd-tU}
ensures that $\sq{U^+(g_n)}$ is bounded in $\lone$. Therefore,
Corollary~A.1.1 in~\cite{Pha09} produces a subsequence, still labeled
$\sq{U^+(g_n)}$, $\eps>0$, and a pairwise disjoint sequence of events
$\{A_n\}_{n\in\N}$ such that
\[
\EE[ U^+(g_n) {\mathbf1}_{{A_n}}]\geq2 \eps>0
\qquad\mbox{for all }n\in\N.
\]
The monotone convergence theorem allows us to exchange $\eps$
in utility for boundedness and obtain the existence of a sequence
$\{r_n\}_{n\in\N}\subseteq\linf_+\cap\sC(x)$ such that $r_n\leq
g_n$ and
$\EE[
U^+(r_n){\mathbf1}_{{A_n}}]\geq\eps$, for all $n\in\N$. Let the sequence
$\{h_n\}_{n\in\N}$ of bounded random variables be defined by
\[
h_n:=x' +
\sum_{k=1}^n r_k {\mathbf1}_{{A_k}}\in x'+\linf_+\subseteq x_0+\linf_+.
\]
For
$\QQ\in\sP$, we have $\langle\QQ,h_n-x'-nx \rangle=\sum_{k=1}^n
\langle\QQ,r_k{\mathbf1}_{{A_k}}-x \rangle\leq n \ac(\QQ)$, so
that $\frac
{1}{n}h_n\in
\sC(x+\frac1n x')\subseteq\sC(x+ x')$ for all $n\in\N$. On the other
hand, since $U(h_n)=U^+(h_n)$, we have
\[
\EE[U(h_n)]\ge\sum_{k=1}^n \EE[
U^+(r_k){\mathbf1}_{{A_k}}]\geq n \eps.
\]
Using (\ref{equaaa}) with $n=2^m$ for
$m\in\N$ produces
\begin{eqnarray*}
2^m \eps
&\leq&
\EE[ U(h_{2^m})] \leq\EE[ U(2^m x' \vee h_{2^m} )] \leq c^m \EE\biggl[
U\biggl(x' \vee\frac{1}{2^m} h_{2^m}\biggr)\biggr]\\
&\leq&
c^m \EE\biggl[U\biggl(x'+\frac{1}{2^m}
h_{2^m}\biggr)\biggr]\leq c^m u(x+2x'),
\end{eqnarray*}
which, thanks
to the fact that $c<2$, implies that $u(x+2x')\geq\eps\lim_m
(2/c)^m=\infty$, a statement in contradiction with the fact that $u$ is
$[-\infty,\infty)$-valued everywhere.
\end{pf}
%
\begin{proposition} \label{prousc} Under Assumptions~\ref{assC} and
\ref{assU}, the primal value function $u$ is upper-semicontinuous.
\end{proposition}
\begin{pf} Thanks to $u$'s concavity and monotonicity, it will be
enough to
show that $u(\ux)\geq\lim_n u(x_n)$ for each sequence $x_n\searrow
\ux=\inf\{x\in\R\dvtx u(x)>-\infty\}$ with $x_n>\ux$. We pick
such a
sequence $\{x_n\}_{n\in\N}$ and use Proposition \ref
{propprimalexistence} to
construct a sequence $\{f_n\}_{n\in\N}$ of random variables such that
$f_n\in
\sC(x_n)$ and $u(x_n)=\UU(f_n)$. By the same argument as in the first
paragraph of the proof of Proposition~\ref{propprimalexistence}, we can
construct a limit $g\in\cap_n \sC(x_n)$ of a sequence of forward convex
combinations, that is, $g_n:= \sum_k \alpha_k^n f_k$ for positive
constants $\alpha_k^n$ summing (over $k$) to one. By Fatou's lemma and
Corollary~\ref{corhedge}(2), we have for $\QQ\in\sP$,
\[
\langle\QQ,g\rangle\le\liminf_{n\to\infty} \sum_k \alpha_k^n
\langle
\QQ,f_k\rangle\le\liminf_{n\to\infty} \sum_k \alpha_k^n \bigl( x_k
+\alpha_\sC(\QQ)\bigr) = \ux+\alpha_\sC(\QQ),
\]
since $x_n \searrow\ux$. Corollary~\ref{corhedge}(2) implies that
$g\in\sC(\ux)$, and so
$u(\ux)\geq\UU(g)$. Using the ideas of the second paragraph of the proof
of Proposition~\ref{propprimalexistence}, we can establish the uniform
integrability of the sequence $\sq{U^+(g_n)}$, and conclude that
$u(\ux)\geq\UU(g) \geq\lim_n u(x_n)$.
\end{pf}
%
\begin{remark}
The upper-semicontinuity of the value function of a utility maximization
problem has been established in the
dissertation~\cite{Sio10} of Siorpaes, in the setting of
utility maximization with random endowment of~\cite{HugKra04} and
applies jointly to the initial wealth $x$ and the initial quantity of the
random endowment. The proof of Proposition~\ref{prousc} uses similar ideas
and generalizes the results of Siorpaes to constrained markets, but
considers only
the initial-wealth variable $x$.
\end{remark}

\subsection{No need to relax $v$}
We start with an observation about continuity of
the upper-hedging-price map.
%
\begin{lemma}
\label{lemuhp-lsc}
Under the Assumption~\ref{assC},
the upper-hedging-price map,
\[
\linf\ni f\mapsto\rho(f):=\inf\{x\in\R\dvtx f\in x+\sC\}\vadjust{\goodbreak}
\]
is
convex, proper and lower $\sigma(\linf,\lone)$-semicontinuous. Moreover,
there exist a
constant $M>0$ such that
%
\begin{equation}
\label{equbd-rho}
|\rho(f)| \leq M+{\|f\|}\qquad\mbox{for all
}f\in\linf.
\end{equation}
\end{lemma}
\begin{pf}
Thanks to Assumption
\ref{assC}, there exists a constant $M>0$ such that $\sC$ contains
the set $-M-\linf_+$. Therefore, $\rho(f)\leq
{\|f\|}+M$, for any $f\in\linf$. To obtain the full bound
(\ref{equbd-rho}), we assume, to the contrary, that there exists a
sequence $\{f_n\}_{n\in\N}$ in $\linf$ such that
\[
\rho(f_n) < -
{\|f_n\|}-n\qquad\mbox{for all }n\in\N.
\]
Therefore, $
f_n+{\|f_n\|}+n\in\CC$ for each $n\in\N$, and, consequently,
$n\in\sC$,
for each $n\in\N$. This is, however, in contradiction with Assumption
\ref{assC}.

Since properness of $\rho$ follows from the bounds in
(\ref{equbd-rho}), and convexity follows directly from the
definition, it remains to show that $\rho$ is
$\sigma(\linf,\lone)$-lower semicontinuous, that is, that its epigraph
\[
\epi\rho=\{(f,x)\in\linf\times\R\dvtx\rho(f)\leq x\}
\]
is
closed. This follows from the fact that $\epi\rho=\{(f,x) \dvtx
f-x\in\sC\}$ is the inverse image of the closed set $\sC$ under the
continuous map $(f,x)\mapsto f-x$ from $\linf\times\R$ to $\linf$.
\end{pf}
%
\begin{lemma} \label{lemZalinescu} Under Assumption~\ref{assC}, for
each $y\geq0$,
we have
\[
\inf_{\QQ\in\sbayp} \ac(\QQ)=\inf_{\QQ\in\slyp} \ac(\QQ).
\]
\end{lemma}
\begin{pf}
For simplicity, we assume that $y=1$.
The set $\sbap$ is $\sigma(\ba,\linf)$-compact by the
Banach--Alaoglu theorem, so we can use the minimax theorem to conclude that
%
\begin{equation}%
\label{eqummax}
\inf_{\QQ\in\sbap} \ac(\QQ) = \inf_{\QQ\in\sbap} \sup_{f\in
\sC}
\langle\QQ,f \rangle= \sup_{f\in\sC} \inf_{\QQ\in\sbap}
\langle\QQ,f \rangle=
\sup_{f\in\sC} \essinf f.\hspace*{-26pt}
\end{equation}
Now we focus on $\inf_{\QQ\in\slp} \ac(\QQ)$. Since
$\ac(\QQ)=\infty$, for $\QQ\notin\lone\setminus\lone_+$,
we have
\[
\inf_{\QQ\in\slp} \ac(\QQ)=\inf_{\QQ\in\sS} \ac(\QQ),
\]
where
$\sS:=\{\QQ\in\lone\dvtx\langle\QQ,1 \rangle=1\}$. Throughout
the rest
of this
proof, we work with the duality between the spaces $\linf$ and
$\lone$, and all notions of continuity and conjugation should be
understood with respect to this duality and the corresponding
weak-$\ast
$ and weak topologies.

We define the map $\gamma\dvtx\linf\to\Rinf$ by
\[
\gamma(f):=
\cases{
x, &\quad $f=x$, a.s., for $x\in\R$,\cr
+\infty, &\quad otherwise.}
\]
The convex conjugate $\gamma^*$ of $\gamma$ is the indicator ${\chi
}_{\sS}$ of $\sS$.
\[
\gamma^*(\QQ):=\sup_{f\in\linf} \bigl( \langle\QQ,f \rangle-
\gamma(f)
\bigr) =
\sup_{x\in\R} x( \langle\QQ,1 \rangle-1 ) = {\chi
}_{\sS}(\QQ
), \qquad\QQ \in\lone.
\]
Next, we define the infimal convolution ${\chi}_{\sC}\Box\gamma$ of
${\chi}_{\sC}$ and
$\gamma$ by
\[
({\chi}_{\sC}\Box\gamma)(f):= \inf_{ g\in\linf} \bigl(
{\chi}_{\sC}(f-g)+\gamma(g) \bigr),\qquad f\in\linf.
\]
Since $\gamma$ is only finite on constants, we have
\[
({\chi}_{\sC}\Box\gamma)(f)=\inf_{x\in\R} \bigl( {\chi}_{\sC}(f-x)+x
\bigr)=\inf\{x\in\R\dvtx f\in x+\CC\}=\rho(f),\qquad f\in\linf.
\]
It follows from Lemma~\ref{lemuhp-lsc} that ${\chi}_{\sC}\Box
\gamma$ is convex, proper and lsc. Consequently, we have
\[
({\chi}_{\sC}\Box\gamma)^{**}(0)= ({\chi}_{\sC}\Box\gamma
)(0)=\rho
(0)=-\sup_{h\in\sC}
\essinf h.
\]
%
%
On the other hand, by Theorem 2.3.1(ix), page 76, in~\cite{Zal02}, we have
$({\chi}_{\sC}\Box\gamma)^* = {\chi}_{\sC}^*+\gamma^*= \ac
+{\chi}_{\sS}$, and so
\begin{eqnarray*}
({\chi}_{\sC}\Box\gamma)^{**}(0) &=& \sup_{\QQ\in\lone} \bigl(
\langle\QQ,0 \rangle-({\chi}_{\sC}\Box\gamma)^*(\QQ) \bigr)\\
&=& \sup_{\QQ
\in\lone}
-\bigl(\ac(\QQ) +{\chi}_{\sS}(\QQ) \bigr)=-\inf_{\QQ\in\sS}
\ac
(\QQ).
\end{eqnarray*}

A comparison with (\ref{eqummax}) yields the statement.
\end{pf}

To prove Lemma~\ref{leminfs-equal}, we need a result from
\cite{KraSch99}. We state a rephrased version whose proof can be read
off the proof of Proposition 3.2, page 924, of~\cite{KraSch99} (in
particular, no additional smoothness assumptions on $V$ are required).
%
\begin{lemma}[(Kramkov and Schachermayer)] \label{labKraSch}
Let $\MM\subseteq\DD$ be
bounded subsets of $\lone_+$ such that:
\begin{longlist}[(3)]
\item[(1)] the mapping $\DD\ni h \to\EE[V(h)]$
attains its minimum at some $\hat{h}\in\DD$;
\item[(2)] $\MM$ is closed under countable convex
combinations;
\item[(3)] there exists a sequence $\{h_n\}_{n\in\N}\subseteq\MM$
which converges to $\hat{h}$ in probability.
\end{longlist}
Then, under Assumption
\ref{assU}, we have
$\inf_{h\in\DD} \EE[V(h)]=\inf_{h\in\MM} \EE[V(h)]$.
\end{lemma}
%
\begin{lemma}
\label{leminfs-equal}
Under Assumptions~\ref{assC} and
\ref{assU}, let $S\subseteq\sbayp$ be of the form
$S=\{\QQ\in\sbayp\dvtx\ac(\QQ)\leq M\}$, for some constant
$M\in\R$. Then, provided that $S\cap\lone\ne\varnothing$, we have
%
\begin{equation}
\label{equinfs-equal}
\inf_{\QQ\in S} \VV(\QQ)= \inf_{\QQ\in S\cap\lone} \VV(\QQ).
\end{equation}
\end{lemma}
\begin{pf}
To simplify the notation, we assume that $y=1$---the general case is
completely analogous. Let $\DD$ denote the set of all
(Radon--Nikodym derivatives of) regular parts of the elements in $S$,
and let $\MM\subseteq\DD$ denote the set of all (Radon--Nikodym
derivatives of) elements of $S\cap\lone$. Since the passage to the
regular part does not increase the total mass, $\DD$ is clearly
bounded in $\lone$.

The statement will follow from Lemma~\ref{labKraSch}, once its
assumptions are verified:\vspace*{8pt}

(1) The set $S$ is a weak-$\ast$ closed (and therefore
compact) subset of $\sbap$, and $\VV$ is lower semicontinuous, so
there exists $\hQ\in\sbap$ at which the infimum on the
left-hand side expression of (\ref{equinfs-equal}) is achieved. It
follows from representation (\ref{equV-reg}) that
${\hat{h}}\in\argmin_{h\in\DD} \EE[V(h)]$, where ${\hat
{h}}:=\frac{d\hQ^r}{d\PP}$.

(2) Let $\{\QQ_n\}_{n\in\N}$ be a sequence of (countably additive)
probability measures in~$\MM$, and let $\{\alpha_n\}_{n\in\N}$ be a sequence
of positive constants with $\sum_n \alpha_n=1$. To show that the
probability measure $\QQ=\sum_n \alpha_n \QQ_n$ belongs to $\MM$, we
need to show that $\ac(\QQ)\leq M$, that is, that $\langle\QQ,f
\rangle\leq M$,
for all $f\in\CC$. This follows by aggregation (combined with monotone
convergence) of the
inequalities $\langle\alpha_n\QQ_n,f \rangle\leq\alpha_n M$ over
$n\in\N$.

(3) We first establish an auxiliary claim. We remind the
reader that for $A\subseteq\lzer_+$, $A^{\circ}$ denotes the polar
of $A$, that is, $A^{\circ}:=\{g\in\lzer_+\dvtx\EE[fg]\leq
1$, for all $f\in A\}$.
%
\begin{claim}
\label{claone}
For $\QQ\in S$, we have
$\QQ^r\in\MM^{\circ\circ}$.
\end{claim}
\begin{pf}
Let us first note that
%
\begin{equation}
\label{equSeq}
S= \{\QQ\in\sbap\dvtx\alpha_{\mathcal{C}'}(\QQ) \le0\},
\end{equation}
where
$\mathcal{C}'\subset\mathbb{L}^\infty$ denotes the weak-$\ast$
closed convex cone generated by $\mathcal{C}-M-
\mathbb{L}^\infty_+$. The inclusion $\supseteq$ clearly holds, and
for the opposite one it suffices to note that $\langle\QQ,\gamma
(f-M-k) \rangle\leq
0$, for all $\QQ\in S$ and all $\gamma\geq0$, $f\in\mathcal{C}$
and $k\in\linf_+$.

By (\ref{equSeq}) we have
\[
\langle\QQ^r,g+1 \rangle\le\langle\QQ,g+1 \rangle=\langle\QQ
,g \rangle+1\le1
\]
for all $\QQ\in S$ and $g\in\CC'$ with $1+g\in\linf_+$. Therefore,
$\QQ^r\in A^{\circ}$, for all $\QQ\in S$, where
$A=(\mathcal{C}'+1)\cap\mathbb{L}^0_+$. Consequently, Claim\vadjust{\goodbreak}
\ref{claone} will be proven once we show that
\[
A^{\circ} \subset
\{S^{\mathbb{L}^1}_+(1)\dvtx\alpha_{\mathcal{C'}}(\QQ) \le0\}
^{\circ\circ}.
\]
To this end we argue by
contradiction and assume that there exists
\[
\hat{\QQ}\in A^{\circ} \setminus\{\QQ\in
S^{\mathbb{L}^1}_+(1)\dvtx\alpha_{\mathcal{C'}}(\QQ) \le
0\}^{\circ\circ}.
\]
In other words, we assume that there exist
$\hat{\QQ}\in A^{\circ}$ and $\hat{h}\in\{\QQ\in
S^{\mathbb{L}^1}_+(1)\dvtx\break\alpha_{\mathcal{C'}}(\QQ) \le0\}^{\circ}$ such
that
%
\begin{equation}\label{eqbipolar1}
\langle\hat{\QQ},\hat{h} \rangle>1,\qquad \langle\hat{\QQ},f
\rangle\le1\qquad
\mbox{for all } f\in A.
\end{equation}
General solidity of polars and the monotone convergence theorem imply
that for all $n\in\N$, we have $\hat{h}\wedge n\in\{\QQ\in
S^{\mathbb{L}^1}_+(1)\dvtx\alpha_{\mathcal{C'}}(\QQ) \le0\}^{\circ}$ and,
for large enough $n\in\N$, it additionally holds that
$\langle\hat{\QQ},\hat{h}\wedge n \rangle>1$. Therefore, we may
assume that
already $\hat{h}\in\linf_+$.

Trivially, (\ref{eqbipolar1}) shows $\hat{h}\notin
A$, and, equivalently, $\hat{h}-1\notin\mathcal{C}'$. By the Hahn--Banach
separation theorem, there exists $\tilde{\QQ}\in\lone$ and $\beta
\in\R$
such that
%
\begin{equation}\label{eqbipolar2}
\langle\tilde{\QQ},\hat{h}-1 \rangle>\beta\ge\langle\tilde
{\QQ},g \rangle
\qquad\mbox
{for all } g\in\mathcal{C}'.
\end{equation}
Given that $\mathcal{C}'$ contains $M'-\linf_+$ for some $M'\in\R$, we
must have $\tilde{\QQ}\in\lone_+$, and since $\mathcal{C}'$ is a cone,
we must also have $\beta= 0$. Nontriviality of $\tilde{\QQ}$ allows
us safely to assume---by scaling, if necessary---that $\|\tilde{\QQ
}\|_1 = 1$.
The second inequality in (\ref{eqbipolar2}) shows $\tilde{\QQ} \in
\{\QQ\in S^{\mathbb{L}^1}_+(1)\dvtx\alpha_{\mathcal{C'}}(\QQ) \le
0\}$. However, we have assumed that $\hat{h} \in\{\QQ\in
S^{\mathbb{L}^1}_+(1)\dvtx\alpha_{\mathcal{C'}}(\QQ) \le0\}^{\circ}$ which
implies $\langle\tilde{\QQ},\hat{h} \rangle\le1$ and thereby
contradicts the
first inequality in (\ref{eqbipolar2}).
\end{pf}

Returning to the proof of (3), we note that the weak-$\ast$ compactness
of $S$ (via the Banach--Alaoglu theorem) guarantees the existence of a
minimizer $\hat{\QQ} \in S$ for the left-hand side of
(\ref{equinfs-equal}). Thanks\vspace*{1pt} to representation
(\ref{equV-reg}), all we need to do is construct a sequence
$\sq{\frac{d\QQ_n}{d\PP}}\subset\MM$ which\vspace*{-1pt} converges
almost surely to the regular part $\frac{d\hat{\QQ}^r}{d\PP}\in\DD$,
and for that we will use a variant of an argument in
\cite{CviSchWan01}. By the bipolar theorem (see~\cite{BraSch99}),
$\MM^{\circ\circ}$ is the closure in probability of the solid hull of
$\MM$. Therefore, there exist sequences
$\{f_n\}_{n\in\N}\subseteq\lzer_+$ and $\{\QQ_n\}_{n\in\N}\subseteq\MM$
such that $\PP$-a.s.
\[
0\le f_n \le\frac{d\QQ_n}{d\PP},\qquad
f_n \to\frac{d\hat{\QQ}^r}{d\PP
} \qquad\mbox{in probability as } n\to\infty.
\]
Furthermore, by passing to a subsequence, $\PP$-a.s. convergence can
be substituted for the convergence in probability. Koml\'os's lemma can
be used to justify the existence of a nonnegative random variable $Y$
and a double array $\{\beta_n^k\dvtx n\in\N, k=1,\ldots,K(n)\}$ with
$0\leq\beta_n^k\leq1$ such that
\[
\sum_{k=n}^{K(n)}\beta_n^k =1,\qquad n\in\N,\qquad
\frac{d\tilde{\QQ}_n}{d\PP} = \sum_{k=n}^{K(n)}\beta_n^k\,
\frac{d\QQ_k}{d\PP} \to Y, \qquad\PP\mbox{-a.s. as }n\to\infty.
\]
It follows from the
convergence $f_n \to\frac{d\hat{\QQ}^r}{d\PP}$ that
\[
\frac{d\hat{\QQ}^r}{d\PP} = \lim_n\sum_{k=n}^{K(n)}\beta_n^k f_k
\le
\lim_n\sum_{k=n}^{K(n)}\beta_n^k \,\frac{d\QQ_k}{d\PP} = Y.
\]
By the convexity of $\MM$, we have $\frac{d\tilde{\QQ}_n}{d\PP}
\in\MM$,
so it suffices to verify the equality $Y= \frac{d\hat{\QQ}^r}{d\PP}$,
a.s.

Since\vspace*{1pt} $S$ is weak-$\ast$ compact, the sequence
$(\tilde{\QQ}_n)_{n\in\N}\subset S$ must have an accumulation point
$\tilde{\QQ}\in S$, which, by Proposition A.1,\vspace*{1pt} page 271,
in~\cite{CviSchWan01}, must satisfy $\frac{d\tilde{\QQ}^r}{d\PP}=Y$.
Assuming that $\PP[ \frac{d\hat{\QQ}^r}{d\PP} < Y]>0$, representation
(\ref{equV-reg}) produces the contradiction
\[
\inf_{\QQ\in S} \VV(\QQ)= \VV(\hat{\QQ}) =
\EE\biggl[V\biggl(\frac{d\hat{\QQ}^r}{d\PP}\biggr)\biggr] >\EE[V(Y)] =
\EE\biggl[V\biggl(\frac{d\tilde{\QQ}^r}{d\PP}\biggr)\biggr] = \VV(\tilde{\QQ}^*),
\]
where the strict inequality is the consequence of the strict decrease
of $V$ which, in turn, follows from the second part of
(\ref{equRAE}). Therefore, we have $Y=\frac{d\hat{\QQ}^r}{d\PP}$,
$\PP$-a.s., and the proof is complete.
\end{pf}
%
\begin{proposition}\label{prov-vba}
Under Assumptions~\ref{assC} and
\ref{assU}, $\vba=v$.
\end{proposition}
\begin{pf} As we already commented in the paragraph following
(\ref{equV-reg}), the inequality $\vba\leq v$ is immediate. It is,
therefore, enough to prove that $\vba(y)\geq v(y)$ for all $y>0$ with
$\vba(y)<\infty$. We fix\vspace*{1pt} such $y>0$, pick $\eps>0$, and
choose a minimizer $\hQy$ for $\vba(y)$. By Lemma~\ref{lemZalinescu},
the family
\[
\sley:=\sey\cap\lone\quad\mbox{where }\sey:=\bigl\{\QQ\in\sbayp\dvtx
\ac(\QQ)\leq\ac\bigl(\hQy\bigr)+\eps\bigr\}
\]
is nonempty. Then, by Lemma~\ref{leminfs-equal}, we have
\begin{eqnarray*}
\vba(y) &=& \ac\bigl(\hQy\bigr) + \VV\bigl(\hQy\bigr)
\geq\ac\bigl(\hQy\bigr)+ \inf_{\QQ\in\sey} \VV(\QQ)
\\[-2pt]
&=& \inf_{\QQ\in\sley} \bigl( \VV(\QQ)+\ac\bigl(\hQy\bigr) \bigr)
\geq\inf_{\QQ\in\sley} \bigl( \VV(\QQ)+\ac(\QQ) \bigr) - \eps
\\[-2pt]
&\geq& \inf_{\QQ\in\slyp} \bigl( \VV(\QQ)+\ac(\QQ) \bigr)-\eps
=v(y)-\eps.
\end{eqnarray*}
\upqed\end{pf}
\begin{pf*}{Proof of Theorem~\ref{thmmain}}
(1) The properties of the function $u$ follow either directly
from the
definition or, in the case of upper semicontinuity, from Proposition
\ref{prousc}. Convexity and lower semicontinuity of $\vba$ follow from
the representation in part (1) of Proposition~\ref{propconjugates}.
Finally, $v$ and $\vba$ are identical, by Proposition~\ref{prov-vba}.

(2) For $x\in\R$ with $u(x)\in\R$, there clearly exists $f\in\sC$ such
that $x+f \ge 0$, and so $x + \langle\QQ,f\rangle\ge0$ for all
$\QQ\in\sP$. If we take the supremum over $f\in\sC$ followed by the
infumum over $\QQ\in\sP$ in this inequality, we get
$x\ge\sup_{\QQ\in\sP} -\alpha _{\sC} (\QQ)$, and consequently, $
\ux\geq\sup_{\QQ\in\sP} -\ac(\QQ) = -\inf_{\QQ\in\sP} \ac (\QQ)$.\vadjust{\goodbreak}

On the other hand, by Corollary~\ref{corhedge}, for $x>
\sup_{\QQ\in\sP} -\ac(\QQ)$, we can find $\eps>0$
such that $\eps-x \in\sC$. Therefore, $u(x)\geq U(\eps)>-\infty$,
and, so
$x \geq\ux$. The second statement follows from the fact that $u$ is proper
and nondecreasing.

(3) The existence of primal optimizers is proven in Proposition
\ref{propprimalexistence}.

(4) The relation (\ref{equconjugate}) is proven in Proposition
\ref{propconjugates}, part (1) and Proposition~\ref{prov-vba}. The
symmetric relation (\ref{eqvconjugate}) follows directly from
(\ref{equconjugate}) and the upper semicontinuity of $u$.
\end{pf*}

\section{\texorpdfstring{A sufficient condition for Assumption \protect\ref{assC}}
{A sufficient condition for Assumption 2.3}}
\label{secsuff}

The closedness in probability of the sets $\CC(x)$, $x\in\R$, is the
central condition of our main results. It is, however, not immediately
obvious how to test its validity in a given model. Thanks to a recent
result of~\cite{CziSch11}, a much more workable sufficient condition
can be given. We start by recalling that
each ($\R^d$-valued) semimartingale $S$ can be represented in terms
of its \textit{predictable characteristics},
\[
S=S^c+F+ \bigl(x{\mathbf1}_{\{|x|\leq1\}}\bigr)\ast(\mu-\tilde{\mu})+
\bigl(x{\mathbf1}_{\{|x|> 1\}}\bigr)\ast\mu,
\]
where $S^c$ is a continuous
semimartingale, $F$ is a predictable process of finite variation,
$\mu$ is the jump measure of $S$ and $\tilde{\mu}$ is its
compensator. Instead of explaining these terms we refer the
reader to the standard reference~\cite{JacShi03}. Furthermore, it is
well known that there exists a nondecreasing
process $B$, a $\R^d$-valued process $b$, a
nonnegative-definite $\R^{d\times d}$-matrix valued process $c$ and a
L\'evy-measure-valued process $\Gamma$, all predictable, such that
\[
F=b\cdot B,\qquad [S^c,S^c]=c\cdot B\quad\mbox{and}\quad\tilde{\mu}= \Gamma
\cdot B.
\]
The triplet $(b,c,\Gamma)$ is usually referred to as the
\textit{triplet of semimartingale characteristics} of $S$.

It can be shown that the measure $\PP\otimes dB$ is $\sigma$-finite
and can, therefore, be replaced by an equivalent probability measure
on the predictable sets of $\Omega\times[0,T]$, which we denote by
$\PP^S$. We refer the reader to~\cite{CziSch11} for a discussion and
the interpretation of the probability measure $\PP^S$ (this measure is
denoted by $\PP_B$ in~\cite{CziSch11}), as well as for the proof of the
following proposition.
%
\begin{proposition}[(Czichowsky and Schweizer~\cite{CziSch11})]
There exists a predictable process $\{ \Pi^S_t \}_{t\in[0,T]}$, with
values in the orthogonal projections in $\R^d$ with the following
property. For predictable processes $\theta,\vp$ with $\theta$ being
$S$-integrable, the following two statements are equivalent:
\begin{longlist}[(2)]
\item[(1)] $\vp$ is $S$-integrable with $\theta\cdot S$ and $\vp
\cdot S$
indistinguishable, and
\item[(2)] $\Pi^S \theta=\Pi^S \vp$, $\PP^S$-a.e.
\end{longlist}
\end{proposition}

We fix a version of such a $\Pi^S$ and we call it the
\textit{projection on the predictable range of $S$}. One can think of
$\Pi^S \theta$ as the ``relevant'' portion of $\theta$, as far as
stochastic integration with respect to $S$ is concerned. It was shown
in~\cite{CziSch11} that closedness of the set of constrained stochastic
integrals is closely related to the interplay between $\Pi^S$ and the
constraint $\kappa$:
%
\begin{theorem}[(Czichowsky and Schweizer~\cite{CziSch11})]
\label{thmclosed}
Let $\kappa$ and $\AA^\kappa$ be as in Section
\ref{sseconv-const}. Then the set of stochastic integrals
$\{H\cdot S\dvtx H\in\AA^\kappa\}$ is closed with respect to the
semimartingale topology if and only if $\Pi^S_t(\omega)
\kappa_t(\omega)$ is a closed subset of $\R^d$, $\PP^S$-a.e.
\end{theorem}
%
\begin{remark}
Since closedness of the set $\Pi^S_t(\omega) \kappa_t(\omega)$ is
going to play a prominent role in the sequel, let us briefly comment
on its financial interpretation. It states, essentially, that when
the constraints are imposed, one should take into account those
aspects of the portfolio that actually matter for the evolution of
the wealth process;
see Section 3 of~\cite{CziSch11} for a detailed explanation. In most
models of interest, $\Pi^S$ is the identity; that is, there are no
``redundant'' assets, and the closedness condition is automatically
satisfied. For other sufficient conditions, see~\cite{CziSch11}.
Let us mention that closedness is guaranteed for all semimartingales
$S$ when, for example, with probability one, for each
$t\in[0,T]$ one of the following three properties holds:
\begin{longlist}[(3)]
\item[(1)] $\kappa_t(\omega)$ compact,
\item[(2)] $\kappa_t(\omega)$ is polyhedral (i.e., representable as an
intersection of finitely many half-planes) or
\item[(3)] the support map $\R^d\ni\bsx\mapsto\sup_{\bsy\in
\kappa_t(\omega)} \bsx^T \bsy$ of $\kappa_t(\omega)$ is
continuous.
\end{longlist}
\end{remark}

Using Theorem~\ref{thmclosed} as one of the central ingredients, we
can prove the sufficiency of our conditions for convex compactness of
$\CC(x)$ in Proposition~\ref{prosufficient}. For the reader's
convenience, we repeat its statement below:
\begin{propositiontwofive*}
Assumption~\ref{assC} holds if the following three conditions are satisfied:
\begin{enumerate}[(3)]
\item[(1)] $\sA\ne\varnothing$;
\item[(2)] the projection $\Pi_t^S(\omega)\kappa_t(\omega)$ is
closed, for
$\PP^S$-a.e.;
\item[(3)] there exist:
\begin{enumerate}[(a)]
\item[(a)] a probability measure $\QQ\sim\PP$;\vspace*{1pt}

\item[(b)] $\hH\in\AA$ with
$\EE^{\QQ}[ (\hH\cdot S)_T]<\infty$ and $\hH\cdot S$ locally bounded;
\item[(c)] a nondecreasing predictable c\`{a}dl\`{a}g process $\{ A_t \}
_{t\in[0,T]}$, with $A_0=0$,
\end{enumerate}
such that
%
\begin{equation}\label{equsuper-N}
H\cdot S - (\hH\cdot S + A) \mbox{ is a
$\QQ$-supermartingale}\qquad\mbox{for all $H\in\AA$.}
\end{equation}
\end{enumerate}
\end{propositiontwofive*}
\begin{pf}
The condition $\AA\ne\varnothing$ implies that $\CC(x)\ne
\varnothing$ for
some $x\in\R$, so it will be enough to show that $\CC(x)$ is convexly
compact.

First, we show that $\CC(x)$ is closed in probability. Let
$\{f_n\}_{n\in\N}$ be a sequence in $\CC(x)$ with
\[
f_n = x+(H^n\cdot S)_T - g_n \to f\qquad\mbox{in probability,}
\]
where $H^n
\in\sA$ and $g_n\in\mathbb{L}^0_+$, for all $n\in\N$.
By passing to a sequence of convex combinations (justified by
Koml\'os's theorem and the fact that our\vadjust{\goodbreak} constraints are convex) we
can---and will---assume that $g_n =0$, $\PP$-a.s., for all $n\in\N$. It
therefore suffices to find $H\in\sA$ such that \mbox{$(H\cdot S)_T\ge
\lim_{n\to\infty} (H^n\cdot S)_T$}.

Let $\NN$ denote the set of all pairs $(\QQ,A)$ [with $\QQ$
as in (3)(a) and $A$ as in (3)(c)] for which there
exists $\hH$ as in (3)(b) such that (\ref{equsuper-N}) holds.
We fix $(\QQ,A)\in\NN$ so that for each element $V^n$ in the
sequence
\[
V^n= (H^n - \hH)\cdot S,\qquad n\in\N,
\]
the process $V^n-A$ is a $\QQ$-supermartingale. In particular, we have
\begin{eqnarray*}
V^n_t-A_t &\geq& \EE^{\QQ}[ V^n_T-A_T|\FF_t]\\
&=&
\EE^{\QQ}[ (H^n\cdot S)_T|\FF_t] -
\EE^{\QQ}[ A_T+(\hH\cdot S)_T|\FF_t] \\
&\geq&- M_t,
\end{eqnarray*}
where $M_t=\EE^{\QQ}[x+(\hH\cdot S)_T+A_T|\FF_t]$ is a
$\QQ$-martingale.
Indeed, $(\hH\cdot S)_T\in\lone(\QQ)$ by assumption and $A_T\in
\lone
(\QQ)$
because the process $-A= (\hH- \hH)\cdot S-A$ is a $\QQ$-supermartingale.

From the above we conclude that the processes $V^n-A+M-M_0$, $n\in\N$,
are uniformly lower bounded $\QQ$-supermartingales starting at zero.
Therefore, we can use the Koml\'os-type lemma (Lemma 5.2(1), page 14,
in~\cite{FolKra97}) to extract a Fatou-convergent sequence of convex
combinations. By the\vspace*{1pt} convexity of our constraint sets,
these convex combinations are still of the form $(\tilde{H}^n-\hH)\cdot
S - A + M-M_0$ and converge toward a lower bounded
$\QQ$-supermartingale, which we write in the form $V-A+M-M_0$, for some
semimartingale $V$. Using the properties of Fatou-convergence and the
already assumed convergence of the terminal values $(H^n\cdot S)_T$, we
have
\[
V_0\leq0 \quad\mbox{and}\quad V_T = f - x - (\hH\cdot S)_T.
\]

Since the processes $M$ and $A$ are independent of $n$, we also have
Fatou-convergence of $V^n$ toward $V$. It is important to note that
Fatou-convergence is measure-independent (as long as we stay in the same
equivalence class), so that, for each pair $(\QQ',A')\in\NN$, the
$\QQ'$-supermartingale $V^n-A'$ Fatou-converges toward $V-A'$. The
processes $\hH\cdot S$ and $A'$ are locally bounded ($\hH\cdot S$ is by
assumption whereas $A'$ is thanks to predictability and the
c\`{a}dl\`{a}g property)
so all
$V^n-A'$ are locally bounded from below, with the same localization
sequence. It follows that their Fatou limit $V-A'$ is a
locally-bounded-from-below local $\QQ'$-supermartingale for each
$(\QQ',A')\in\NN$.

The next step is to apply a version of the optional decomposition theorem
developed in~\cite{FolKra97}, namely Theorem 3.1 on page 6.
We need to check that all of its assumptions\vspace*{2pt} are
satisfied, that is, that the family $\sS$ of semimartingales
$\sS=\{(H-\hH)\cdot S\dvtx H\in\AA\}$ satisfies:
\begin{longlist}[(4)]
\item[(1)] $\sS$ is predictably convex (in the language of~\cite{FolKra97});
\item[(2)] $\sS$ contains processes locally bounded from below;\vadjust{\goodbreak}
\item[(3)] $\sS$ is closed in the semimartingale topology for
uniformly-bounded from
below sequences (Assumption 3.1 in~\cite{FolKra97});
\item[(4)] $\sS$ contains the constant process $0$.
\end{longlist}
Indeed,\vspace*{1pt} (1) follows from the convexity of $\kappa$, (2)
holds thanks to the local boundedness of $\hH\cdot S$, (3) is the
content of Theorem~\ref{thmclosed} and (4) is true by the construction
of $\sS$.

Therefore, the fact that $V-A$ is a $\QQ$-local supermartingale
for each $(\QQ,A)\in\NN$ and Theorem 3.1 in~\cite{FolKra97} allow us
to conclude that there exists $H\in\A$ such that
\[
V=V_0 + (H-\hH)\cdot S - C
\]
for some nondecreasing, nonnegative, c\'{a}dl\'{a}g, and adapted process $C$.
We then have the representation
\[
f = x+V_T + (\hH\cdot S)_T = x+ (H \cdot S)_T +
V_0 - C_T \le x+ (H \cdot S)_T.
\]

To finish the proof we need to show that $\CC(x)$ is bounded in
probability. For $f\in\CC(x)$ we let $H\in\AA$ be such that
$x+H\cdot S \geq f$ and pick $(\QQ,A)\in\NN$. Then
\[
\EE^{\QQ}[f]\leq x+ \EE^{\QQ}\bigl[ \bigl((H-\hH)\cdot S\bigr)_T - A_T\bigr] +
\EE^{\QQ}[ (\hH\cdot S)_T+ A_T]\leq M_0,
\]
where---as before---$M_0=x+\EE^{\QQ}[ (\hH\cdot S)_T+ A_T]<\infty$.
This shows that
$\CC(x)$ is bounded in $\lone(\QQ)$, so, by Markov's inequality, it
is bounded in probability under $\QQ$ and, by equivalence, also under
$\PP$.
\end{pf}
%



\printaddresses

\end{document}